\documentclass[12pt]{iopart}
\usepackage{graphicx}
\usepackage{epsfig}
\usepackage{iopams}

\begin{document}

\title{Quantum dissonance provide power to deterministic quantum computation with single qubit}

\author{Mazhar Ali}
\address{Naturwissenschaftlich-Technische Fakult\"at, Universit\"{a}t Siegen, Walter-Flex-Stra\ss e 3, 57068 Siegen, Germany\\
Department of Electrical Engineering, COMSATS Institute of Information Technology, 22060 Abbottabad, Pakistan}
\ead{mazharaliawan@yahoo.com}

\begin{abstract}
Mixed state quantum computation can perform certain tasks which are believed to be efficiently intractable on a classical computer. For a 
specific model of mixed state quantum computation, namely, {\it deterministic quantum computation with a single qubit} (DQC1), recent 
investigations suggest that quantum correlations other than entanglement might be responsible for the power of DQC1 model. However, strictly 
speaking, the role of entanglement in this model of computation was not entirely clear. We provide conclusive evidence that there are instances 
where quantum entanglement is not present in any part of this model, nevertheless we have advantage over 
classical computation. This establishes the fact that quantum dissonance (quantum correlations) present in fully separable states provide power to 
DQC1 model. 
\end{abstract}


\maketitle

\section{Introduction}

Quantum computing promises to solve certain problems in a polynomial time which is much faster than the algorithms designed 
for classical computers \cite{Shor-Grover-1997}. This discovery has attracted tremendous interests and efforts to exploit quantum mechanics for 
information processing tasks. In addition, this research may offer several other types of quantum technologies. There are several fundamental 
questions of interest in this domain which need further attention. One of the main issue is to identify the problems which are hard to solve with 
classical resources. Having this identification, one could ask whether these problems can be solved efficiently on a 
quantum computer or not. If the answer to this question is positive, then it becomes crucial to investigate the resources providing this
power to quantum computers. One apparent feature for this power is frequently attributed to quantum parallelism, which is based on interference 
phenomena derived from the superposition principle. This unique feature of quantum mechanics can create entangled states among more than one 
particle. The role of entanglement in quantum algorithms and other information processing tasks is under investigation. It was shown that 
multipartite entanglement must grow unbounded with the system size if a pure-state quantum computation is to attain an exponential speedup over 
its classical counterpart \cite{Jozsa-Linden-PRSL-2003}. There are instances where entangled states are required to perform a task which is simply 
not possible in classical domain \cite{Horodecki-RMP-2009,NC-QIQC-2000}. However, entanglement is not the only type of correlation useful for 
quantum technology and there are some other quantum correlations which also offer some advantage 
\cite{Bennett-PRA59-1999, Braunstein-PRL83-1999, Horodecki-PRA71-2005, Niset-PRA74-2006}. Meyer \cite{Meyer-PRL85-2000} discovered a quantum search 
algorithm that uses no entanglement. There are also some oracle-based problems that can be solved without entanglement, nevertheless with advantage 
over best known classical algorithms \cite{Biham-Kenigsberg-2006}. 

Although pure state quantum computation has several successful experimental implementations, nevertheless one problem with these schemes is the 
scalability issue and another main problem is to tackle decoherence. In general any quantum computer may interact with 
environment leading to non unitary evolution and its hard to start with pure states which get mixed due to decoherence. Quantum error correction 
techniques tackle this problem, however one may consider the situation where there are no errors or interactions with environment, but the initial 
state is highly mixed. This gives rise to mixed-state quantum computation introduced by Knill and Laflamme \cite{Knill-Laflamme-PRL81-1998}. 
Although the power of mixed-state quantum computation is strictly less than the power of pure-state quantum computation, nevertheless there are 
tasks which are seemingly impossible to achieve with classical computers but they can be achieved with a mixed-state quantum computer 
\cite{Datta-PRL100-2008, Datta-PRA72-2005, Datta-PhDThesis, Datta-IJQI9-2011}. One specific model of computation is 
deterministic quantum computation with one qubit (DQC1). This model works by considering collection of qubits in the completely mixed state as a 
quantum register, coupled to a single control qubit that has some nonzero purity. This device is known to provide an exponential speedup over the 
best known classical algorithm for estimating the normalized trace of a unitary matrix. 
It was shown that quantum discord is responsible for the speedup in DQC1 \cite{Laflamme-QIC2-2002}.  
Datta and coworkers \cite{Datta-PRL100-2008, Datta-PRA72-2005, Datta-PhDThesis, Datta-IJQI9-2011} have investigated the role of entanglement in 
DQC1 and found that the model may have some limited amount of entanglement which does not increase with the system size. Moreover, for certain 
parameter range the final state has a positive partial transpose (PPT) meaning no distillable entanglement if there is some (that is, PPT region might 
contain some bound entanglement), nevertheless the model retains its exponential advantage. With these findings, they conjectured that 
entanglement may not be responsible for the speedup of DQC1 and some other non-classical correlations supply power to this model. These findings 
are interesting and convincing, however strictly speaking, the role of entanglement in this model needs further attention. The previous studies 
have neither proved nor refuted the existence of bound entanglement in DQC1 model. If there exist some bound entanglement in the region of quantum 
advantage then one could not attribute the power of QDC1 solely to quantum correlations other than entanglement. There are some related studies where 
mixed states are used in quantum computing, for instance, in the parallel quantum computing in ensemble quantum computing 
\cite{Xiao-PRA66-2002, Long-PRA69-2004}, and duality quantum computing \cite{Long-CTP45-2006} where 
no or little entanglement are responsible for the speedup of computing over classical counterparts.

Recently, the existence of non-classical correlations using four qubit DQC1 model of computation has been reported \cite{Passante-PRA84-2011}. 
Even in this experiment, the existence or absence of entanglement is not much clear. Two related experiments to witness non-classical correlations 
in DQC1 setup has also been reported \cite{Auccaise-PRL107-2011, Lanyon-PRL101-2008}. 
With growing interest in DQC1 model, it becomes utmost important to investigate the instances where one can show explicitly the absence or presence 
of entanglement while having quantum advantage. As a concrete example, we consider the case of simplest multipartite system of three qubits to 
demonstrate that DQC1 can operate with quantum advantage having no entanglement in any part of model. We have found that for certain instance of 
typical unitaries \cite{Datta-PhDThesis}, PPT region of the final state is always fully separable. Hence, one can attribute the power 
of DQC1 to non-classical correlations (quantum dissonance) present in fully separable states. 

This paper is organized as follows. In Section \ref{Correlations}, we briefly describe classical and quantum correlations both in bipartite and 
multipartite quantum systems. In Section \ref{DQC1-model}, we review DQC1 model and describe an example where this problem can be solved for three 
qubits. In Section \ref{relation}, we provide a brief review of multipartite entanglement and the discussion of entanglement in example studied in 
previous section. Finally, we offer some conclusions in Section \ref{conclusion}.

\section{Quantum correlations other than entanglement}\label{Correlations}

Recent investigations suggest that quantum correlations and quantum entanglement are not always equivalent resources. Quantum correlation is a 
general term which might describe entanglement but there are some other non-classical correlations which can even exist in 
separable states. For bipartite systems, such non-classical correlations are called quantum discord \cite{Ollivier-PRL88-2001, Henderson-JPA34-2001}. 
Quantum discord is measured by the difference of two classically equivalent formulations of mutual information. Quantum mutual information is an 
information-theoretic measure of the total correlation in a bipartite quantum state \cite{Groisman-PRA72-2005}. In particular, if $\rho^{AB}$ denotes 
the density operator of a composite bipartite system $AB$, and $\rho^A$ ($\rho^B$) the density operator of part $A$ ($B$), respectively, then the 
quantum mutual information is defined as
\begin{eqnarray}
\mathcal{I} (\rho^{AB}) = S (\rho^A) + S (\rho^B) - S(\rho^{AB})\, , \label{Eq:QMI}
\end{eqnarray}
where $S(\rho) = - \mathrm{tr} \, ( \rho \, \log_2 \rho )$ is the von Neumann entropy. It was suggested that the total correlations in a 
given quantum state can be split into two parts, that is as a sum of classical correlation $\mathcal{C}(\rho^{AB})$ and quantum correlation 
$\mathcal{Q} (\rho^{AB})$ \cite{Ollivier-PRL88-2001, Henderson-JPA34-2001}, that is
\begin{eqnarray}
\mathcal{I} (\rho^{AB}) = \mathcal{C} (\rho^{AB}) + \mathcal{Q} (\rho^{AB}) \, . 
\end{eqnarray}
In order to quantify quantum discord, Ollivier and Zurek \cite{Ollivier-PRL88-2001} suggested 
the use of von Neumann type measurements. Let the projection operators $\{ B_k\}$ describe a von Neumann measurement for subsystem $B$ only, then 
the conditional density operator $\rho_k$ associated with the measurement result $k$ is 
\begin{eqnarray}
\rho_k = \frac{1}{p_k} \, (I \otimes B_k) \, \rho \, (I \otimes B_k) \,,
\end{eqnarray}
where the probability $p_k = \mathrm{tr} [(I \otimes B_k) \, \rho \, (I \otimes B_k)]$. The quantum conditional entropy with respect to this 
measurement is given as \cite{Luo-PRA77-2008} 
\begin{eqnarray}
S (\rho | \{B_k\}) := \sum_k p_k \, S(\rho_k) \, ,
\label{Eq:QCE}
\end{eqnarray}
and the associated quantum mutual information of this measurement is defined as
\begin{eqnarray}
\mathcal{I} (\rho|\{B_k\}) := S (\rho^A) - S(\rho|\{B_k\}) \, . \label{Eq:QMIM} 
\end{eqnarray}
A measure of the resulting classical correlations is provided \cite{Ollivier-PRL88-2001, Luo-PRA77-2008} by 
\begin{eqnarray}
\mathcal{C}_B (\rho) := \sup_{\{B_k\}} \, \mathcal{I} (\rho|\{B_k\}) \, , \label{Eq:CC} 
\end{eqnarray}
and finally one can obtain quantum discord as $ \mathcal{Q}_B (\rho) := \mathcal{I}(\rho) - \mathcal{C}_B (\rho)$. Alternative derivations can be 
performed for $\mathcal{Q}_A (\rho)$ and in general $\mathcal{Q}_A (\rho) \neq \mathcal{Q}_B (\rho)$ \cite{Bylicka-PRA81-2010}. 
Due to complicated extremization procedure, the evaluation of quantum discord has been done for specific states for bipartite systems  
\cite{Mali-work, Sai-JPA45-2012, Galve-PRA83-2011, Lu-PRA83-2011, Al-Qasimi-PRA83-2011, Girolami-PRA83-2011, Chen-PRA84-2011, Vedral-et-al}. 

\begin{figure}
\scalebox{1.75}{\includegraphics[width=1.55in]{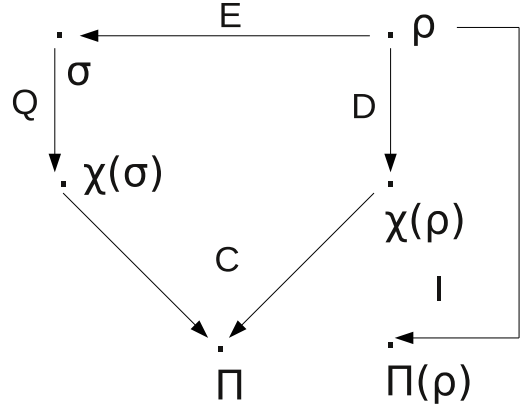}}
\centering
\caption{The geometrical picture of various correlations present in a state. Each type of correlation may be thought of as a distance measure from a 
state to corresponding state.}
\label{Fig:Gpic}
\end{figure}

It was observed that above entropic definition of quantum mutual information and quantum discord can not be generalized for multipartite systems. 
To tackle this problem a geometrical way to compute these correlations has been introduced \cite{Modi-PRL104-2010}. Let $\rho \in \mathcal{E}$ 
(the set of entangled states), $\sigma \in \mathcal{S}$ (the set of separable states), $\chi \in \mathcal{C}$ (the set of classical states), 
and $\pi \in \mathcal{P}$ (the set of product states), then we define
\begin{eqnarray}
I_\rho \, &\equiv& \, S(\rho \, \| \, \pi_\rho) \quad (mutual \, information) \nonumber \\&
E \, =& \min_{\sigma \in \mathcal{S}} \, S(\rho \, \| \, \sigma) \quad (entanglement) \nonumber \\&
D \, =& \min_{\chi \in \mathcal{C}} \, S(\rho \, \| \, \chi) \quad (quantum \, discord) \nonumber \\&
Q \, =& \min_{\chi \in \mathcal{C}} \, S(\sigma \, \| \, \chi) \quad (quantum \,dissonance) \nonumber \\&
C \, =& \min_{\pi \in \mathcal{P}} \, S(\chi \, \| \, \pi) \quad (classical \, correlation), 
\end{eqnarray}
where $S(x \, \| \, y)$ is the relative entropy between two quantum states $x$ and $y$. Although these definitions are mathematically elegant, but 
quite hard to compute for an arbitrary initial state. Figure \ref{Fig:Gpic} depicts these definitions graphically.  

\section{Deterministic quantum computation with a single qubit}\label{DQC1-model}
 
\begin{figure}
\scalebox{1.75}{\includegraphics[width=1.55in]{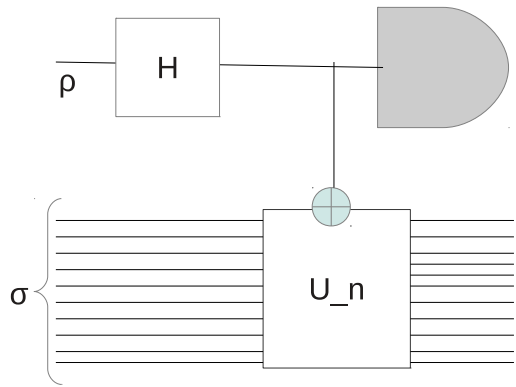}}
\centering
\caption{The Hadamard gate is applied on special qubit $\rho$ before controlled unitary operation on register qubits. Finally, the special qubit
is being measured.}
\label{Fig:DQC1model}
\end{figure}

The original model of mixed-state quantum computing with one qubit proposed by Knill and Laflamme \cite{Knill-Laflamme-PRL81-1998} is usually called 
``power of one qubit`` as it only requires one qubit for measurements. This model has been slightly generalized later 
\cite{Datta-PRL100-2008, Datta-PRA72-2005, Datta-PhDThesis, Datta-IJQI9-2011}. The setup of this scheme is shown in Figure \ref{Fig:DQC1model}. 
The circuit consists of a register of $n$ qubits initially all in maximally mixed state, that is, $\sigma = I_n/2^n$. The special qubit has an 
initial state $\rho = (I + \alpha \, Z)/2$, where $0 \leq \alpha \leq 1$, and $Z$ is the standard Pauli matrix. First a Hadamard gate is applied to 
the special qubit and then a control unitary gate is applied on register qubits only when special qubit is in state $|1\rangle$ otherwise 
simply identity matrix is applied.  
The final state of all qubits after these operations can be compactly written \cite{Datta-PhDThesis} as
\begin{eqnarray}
\rho_{n+1} (\alpha) = \frac{1}{N} \left( \begin{array}{cc}
I_n & \alpha \, U_n^\dagger\\ 
\alpha \, U_n & I_n \end{array}
\right) \,,
\label{Eq:SAO}
\end{eqnarray}
where $N = 2^{n+1}$. Let us now measure the special qubit in $X$ basis, where $X$ being the Pauli matrix, and leave the register qubits untouched. 
The expectation value of these measurements can be written as 
\begin{eqnarray}
\langle X \rangle_{\rho_{n+1}(\alpha)} = {\rm tr} \big[ \, (X \otimes I_n) \, \rho_{n+1}(\alpha) \, \big] = 1/\mathcal{N} \, 
\Re \big[ \, {\rm tr} (U_n) \, \big] \, , 
\end{eqnarray}
where $\mathcal{N}$ is normalization factor. Similarly, measurements in the Pauli matrix $Y$ give expectation value 
\begin{eqnarray}
\langle Y \rangle_{\rho_{n+1}(\alpha)} = {\rm tr} \big[ \, (Y \otimes I_n) \, \rho_{n+1}(\alpha) \, \big] = 1/\mathcal{N} \, 
\Im \big[ \, {\rm tr} (U_n) \, \big] \, . 
\end{eqnarray}
Hence, we can estimate the normalized trace of a unitary \cite{Datta-PhDThesis}. 
This brief description is sufficient to understand this model. Interestingly, several problems like, calculation of fidelity decay in quantum chaos, 
problems in quantum meteorology and estimation of Jones Polynomials from Knot Theory can be reduced to evaluation of normalized traces of 
particularly unitary matrices \cite{Datta-PhDThesis, Datta-arXiv2011}. Therefore, this simple quantum computer could be utilized to study and solve 
problems in several areas.    

Now we ask the question that what are the correlations present in mixed state (\ref{Eq:SAO}) which provide power to DQC1. 
One obvious approach is to look for entanglement in the state and the findings reveal that this question is not easy to 
answer due to various issues \cite{Datta-PhDThesis}. However, it was 
shown that the special qubit is always unentangled with the $n$ unpolarized qubits, no  matter what $U_n$ is used \cite{Paulin}. In addition, the 
marginal state of $n$ unpolarized qubits remains maximally mixed, which means that these qubits are not entangled among 
themselves at any stage of this algorithm. Datta \cite{Datta-PhDThesis} looked for entanglement when the special qubit is grouped among subset of 
unpolarized qubits and used {\it multiplicative negativity} $\mathcal{M}$ as a measure of entanglement. Multiplicative negativity can be defined as
\begin{eqnarray}
\mathcal{M}(\rho) = 1 + \mathcal{N}(\rho) \, ,
\end{eqnarray}
where $\mathcal{N}(\rho)$ is negativity \cite{Vidal-PRA65-2002}. Hence, multiplicative negativity is $"1"$ for PPT states, it can only capture 
entanglement which is distillable and ignores bound entanglement. The amount of entanglement depends on the unitary matrix $U_n$ and on bipartitions. 
In particular, Datta constructed a family of unitaries $U_n$ such that for $\alpha > 1/2$, $\rho_{n+1} (\alpha)$ is entangled for all bipartitions 
that put the special qubit with last unpolarized qubit in different parts. 

To construct $U_n$, consider a two-qubit unitary matrix
\begin{eqnarray}
U_2 \equiv \left( \begin{array}{cc}
A_1 & C_1\\ 
D_1 & B_1 \end{array}
\right) \,,
\label{Eq:un2}
\end{eqnarray}
where $A_1$, $B_1$, $C_1$ and $D_1$ are single qubit matrices that satisfy 
$A_1^\dagger \, A_1 + D_1^\dagger \, D_1 = B_1^\dagger \, B_1 + C_1^\dagger \, C_1 = I_1$ and $A_1^\dagger \, C_1 + D_1^\dagger \, B_1 = 0$ to 
ensure that $U_2$ is unitary. The $n$-qubit unitary $U_n$ is defined in compact notation like Eq.(\ref{Eq:SAO}) as
\begin{eqnarray}
U_n \equiv \left( \begin{array}{cc}
I_{n-2} \otimes A_1 & X_{n-2} \otimes C_1\\ 
X_{n-2} \otimes D_1 & I_{n-2} \otimes B_1 \end{array}
\right) \,.
\label{Eq:unN}
\end{eqnarray} 
It turns out that by performing partial transpose of $\rho_{n+1} (\alpha)$, on the part that does not include the special qubit, 
negativity $\mathcal{M} (\rho_{n+1} (\alpha))$ is equal to negativity $\mathcal{M} (\rho_3 (\alpha))$, 
that is, $\mathcal{M} (\rho_{n+1} (\alpha)) = \mathcal{M} (\rho_3 (\alpha))$ 
where $\rho_3(\alpha)$ is three-qubit mixed state. For a specific choice of $U_2$ given as
\begin{eqnarray}
A_1 = \left( \begin{array}{cc}
0 & 0\\ 
0 & 1 \end{array}
\right) \,, \quad B_1 = \left( \begin{array}{cc}
1 & 0\\ 
0 & 0 \end{array}
\right) \,,\\\nonumber C_1 = \left( \begin{array}{cc}
0 & 1\\ 
0 & 0 \end{array}
\right) \,,\quad D_1 = \left( \begin{array}{cc}
0 & 0\\ 
1 & 0 \end{array}
\right) \,, 
\label{Eq:spun2}
\end{eqnarray}
the three qubit state $\rho_3(\alpha)$ turns out to be
\begin{eqnarray}
\rho_3 (\alpha) = \frac{1}{8} \left( \begin{array}{cccccccc}
1 & 0 & 0 & 0 & 0 & 0 & 0 & \alpha \\ 
0 & 1 & 0 & 0 & 0 & \alpha & 0 & 0 \\
0 & 0 & 1 & 0 & 0 & 0 & \alpha & 0 \\
0 & 0 & 0 & 1 & \alpha & 0 & 0 & 0 \\
0 & 0 & 0 & \alpha & 1 & 0 & 0 & 0 \\
0 & \alpha & 0 & 0 & 0 & 1 & 0 & 0 \\
0 & 0 & \alpha & 0 & 0 & 0 & 1 & 0 \\
\alpha & 0 & 0 & 0 & 0 & 0 & 0 & 1 \end{array}
\right) \,.
\label{Eq:rho3ap}
\end{eqnarray}
The eigenvalues of the partially transposed matrix are given as
\begin{eqnarray}
Sp (\tilde{\rho}_3(\alpha)) = \bigg\{ \frac{1+2 \alpha}{8}, \, \frac{1}{8}, \, \frac{1}{8},\, \frac{1}{8}, \, \frac{1}{8}, 
\, \frac{1}{8}, \,\frac{1}{8}, \, \frac{1-2  \alpha}{8} \, \bigg\} \, .
\end{eqnarray}

Multiplicative negativity for this apparently simple state is 
\begin{eqnarray} 
\mathcal{M} (\rho_{n+1} (\alpha)) = \mathcal{M} (\rho_3 (\alpha)) = \max \bigg[ \, 1 \, , \frac{2 \alpha + 3}{4} \, \bigg] \, ,  
\end{eqnarray}
which has a maximum value of $5/4$ for $\alpha = 1$ and is conjectured to be maximum for all $U_n$ \cite{Datta-PhDThesis}. 
The state $\rho_3(\alpha)$ is PPT under all bipartitions for $0 < \alpha \leq 1/2$ and the entanglement properties of $\rho_3(\alpha)$ in this 
range are not known explicitly. However, it was shown that quantum discord is positive in the range $0 < \alpha \leq 1/2$ across this partition. 
Due to the arguments that as the special qubit is not entangled at any stage with unpolarized qubits but nevertheless 
quantum correlated across this partition,  it was believed that even though if there were any entanglement which might be vanishingly small 
in this region, quantum discord is strictly larger and might provide speed up to DQC1 model. 

We will now demonstrate that PPT part of this state is fully separable, meaning that for $ 0 < \alpha \leq 1/2$, there is no entanglement of any 
kind in the state $\rho_3(\alpha)$, however, there are some quantum correlations present in the state for every range of $\alpha$. If one regards 
this problem as a bipartite system with quantum register as a single quantum system, then these correlations are namely quantum discord. However, 
strictly speaking this is a multipartite system and these non-classical correlations which are identified as quantum dissonance 
provide speed up to DQC1. Before we demonstrate this result, we first briefly review the idea of entanglement in multipartite systems.
   
\section{Quantum entanglement in multipartite systems}\label{relation}

We consider a $N$-partite quantum system with associated Hilbert space $\mathcal{H} = \mathcal{H}_1 \otimes \ldots \otimes \mathcal{H}_N$ 
having dimension $D = k^N$, where $k$ is the dimension of each quantum system. For the sake of simplicity, we assume that each party has the 
same dimension. For our purpose $k = 2$, which corresponds to a qubit. A pure state $|\psi\rangle \in \mathcal{H}$ is called fully separable 
(FS) if it can be written as $|\psi^{fs} \rangle = |\psi_1 \rangle \otimes \ldots \otimes |\psi_N \rangle$. Extending this argument we recognize 
a mixed state $\rho \in \mathcal{H}$ to be fully separable if it can be written as a convex combination of fully separable pure states as
\begin{eqnarray}
\rho^{fs} = \sum_j \, p_j \, |\psi_j^{fs} \rangle\langle \psi_j^{fs} | \, , 
\end{eqnarray}
where $p_j \geq 0$ and $\sum_j \, p_j = 1$. These states do not contain any entanglement. Any multipartite pure state is called biseparable (BS) 
if it is separable under some bipartition. For example $|\Psi^{bs}\rangle = |\Psi_A \rangle \otimes |\Psi_{\bar{A}} \rangle$ with respect to 
some bipartition $A | \bar{A}$ where $A$ denotes some subset of subsystems and $\bar{A}$ its complement. These states might contain some entanglement 
as $\Psi_A$ and/or $\Psi_{\bar{A}}$ might not be separable. We can generalize this definition to multipartite mixed states straight forwardly. A 
multipartite mixed state is called biseparable if it can be written as 
$\rho^{bs} = \sum_j \, p_j \, |\psi_j^{bs}\rangle\langle \psi_j^{bs}|$, where $|\psi_j^{bs}\rangle$ might be biseparable under different 
partitions. Finally a multipartite state is called genuinely entangled (GME) if it is neither fully separable nor biseparable. The description of 
genuine entanglement for multipartite mixed states is quite challenging. This type of entanglement is of particular interest to generate in 
experiments \cite{Monz-PRL106-2011, Leibfried-2004, Roos-2004, Kiesel-2005}. Considerable efforts have been devoted for its characterization, 
quantification and detection \cite{Bastian-PRL106-2011, Sabin-EPJD48-2008, A-Kay, Guehne-NJP12-2010, Guehne-PLA375-2011, Duer-PRA61-2000, 
Acin-PRL87-2001, Toth-PRL102-2009, Osterloh-2006}. 

Another related problem is to check the full separability of an arbitrary initial multipartite mixed state. Even for three qubit system partial 
results have been worked out recently \cite{A-Kay, Guehne-PLA375-2011} for the set of GHZ-diagonal states. We would use these results to analyze 
the full separability of $\rho_3(\alpha)$.  To this end, we note that one could write $\rho_3(\alpha)$ as a convex combination of a GHZ-diagonal 
state $\omega(\alpha)$ and another state $\eta$, that is
\begin{eqnarray}
\rho_3(\alpha) = (1-\frac{\alpha}{2}) \, \omega(\alpha) + \frac{\alpha}{2} \, \eta \, , 
\end{eqnarray}
where 
\begin{eqnarray}
\omega (\alpha) = \frac{1}{8 - 4 \alpha} \times \nonumber\\ \left( \begin{array}{cccccccc}
1 & 0 & 0 & 0 & 0 & 0 & 0 & \alpha \\ 
0 & 1 - \alpha & 0 & 0 & 0 & 0 & 0 & 0 \\
0 & 0 & 1 - \alpha & 0 & 0 & 0 & 0 & 0 \\
0 & 0 & 0 & 1 & \alpha & 0 & 0 & 0 \\
0 & 0 & 0 & \alpha & 1 & 0 & 0 & 0 \\
0 & 0 & 0 & 0 & 0 & 1 - \alpha & 0 & 0 \\
0 & 0 & 0 & 0 & 0 & 0 & 1 - \alpha & 0 \\
\alpha & 0 & 0 & 0 & 0 & 0 & 0 & 1 \end{array}
\right) \,,
\label{Eq:sgma}
\end{eqnarray}
and 
\begin{eqnarray}
\eta = \frac{1}{2} |\phi\rangle\langle \phi| + \frac{1}{2} |\psi\rangle\langle \psi| \, ,   
\end{eqnarray}
with $|\phi\rangle = |+\rangle_A \otimes |0\rangle_B \otimes |1\rangle_C$ and $|\psi\rangle = |+\rangle_A \otimes |1\rangle_B \otimes |0\rangle_C$, 
and $|+\rangle = (| 0 \rangle + | 1\rangle)/\sqrt{2}$. This means that $\eta$ is fully separable state, so now the question is to analyze 
$\omega (\alpha)$. We will show now that in the range $0 < \alpha \leq 1/2$, $\omega (\alpha)$ is also fully separable, hence it follows that 
the convex combination of two fully separable state is again a fully separable state. This would establish the fact that in range 
$0 < \alpha \leq 1/2$, $\rho_3(\alpha)$ does not contain any type of entanglement.

As $\omega (\alpha)$ belongs to the set of GHZ-diagonal states, and the entanglement properties for the set of GHZ-diagonal states are worked out 
recently \cite{ A-Kay, Guehne-PLA375-2011}. For this purpose, we utilize a recent result, which we name as {\it Kay's criterion} \cite{A-Kay}, 
which is a necessary and sufficient criterion for full separability for states which are PPT under each partition. To explain this criterion, 
we prefer to express GHZ-diagonal states in the so called stabilizer formalism. The GHZ-diagonal states for three qubits can be written as
\begin{eqnarray}
\rho =& \frac{1}{8} \bigg[\, III + \lambda_2 \, ZZI + \lambda_3 \, ZIZ + \lambda_4 \, IZZ + \lambda_5 \, XXX \nonumber\\& 
+ \lambda_6 \, YYX + \lambda_7 \, YXY + \lambda_8 \, XYY \, \bigg]\,, 
\end{eqnarray}
where $X$, $Y$, $Z$, denote the Pauli matrices, $\lambda_i$ are real numbers and $I$ is the $2 \times 2$ identity matrix. For simplicity we have 
omitted the tensor product symbols. It has been shown \cite{A-Kay} that if it holds that 
``$\lambda_5 \, \lambda_6 \, \lambda_7 \, \lambda_8 \leq 0$``, that is, this product is negative semi-definitive then the PPT criterion is 
necessary and sufficient condition for full separability. It is simple to find that for $ \omega(\alpha)$, we have  
\begin{eqnarray}
\lambda_5  = - \lambda_8 =  \frac{2 \, \alpha}{2-\alpha} \,, \quad \lambda_6 = \lambda_7 = 0\,, 
\end{eqnarray}
so that $\lambda_5 \, \lambda_6 \, \lambda_7 \, \lambda_8 = 0$. As $\omega (\alpha)$ is PPT for all partitions in the range $0 < \alpha \leq 1/2$, 
hence it follows that it is fully separable. One gets the same conclusion via using the different method described in \cite{Guehne-PLA375-2011}. 
So far, we have shown that PPT region of $\rho_3(\alpha)$ is fully separable.  
       
Our next step is to demonstrate that fully separable region of $\rho_3(\alpha)$ contain non-classical correlations. This may be done by two 
approaches. One direct method is to directly compute the amount of non-classical correlations using any distance measure. However, one has to 
compute the nearest separable or classical state to any given initial state, which is not an easy problem due to complicated optimization. 

\begin{figure}
\scalebox{1.95}{\includegraphics[width=1.75in]{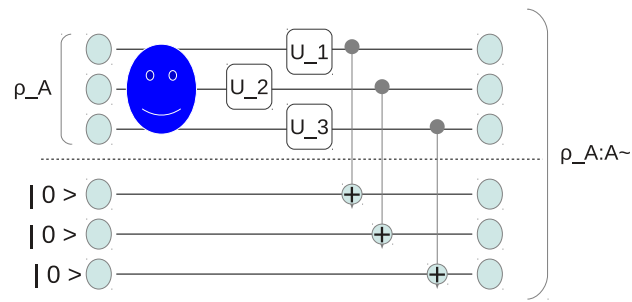}}
\centering
\caption{The activation protocol of non-classical correlations for three qubits. It follows that if there are some non-classical correlations in 
initial state than no matter what unitaries are applied by adversary, there will always be distillable entanglement across the splitting shown.}
\label{Fig:APNCC}
\end{figure}

Another indirect way for existence of non-classical correlations is to activate them into distillable entanglement \cite{Piani-PRL106-2011}. 
See also \cite{Streltsov-PRL106-2011}. In this method $n$ parties (system ${\bf A}$) have $n$ ancilla systems (${\bf \tilde{A}}$) at their disposal 
and the aim is to generate distillable entanglement across ${\bf A}$ and ${\bf \tilde{A}}$ split. An adversary is allowed to perform local unitary 
operations on each subsystems of ${\bf A}$ before $n$ parties perform CNOT gate between their respective system-ancilla pair. It follows that there 
can only be distillable entanglement across ${\bf A}$ and ${\bf \tilde{A}}$ split if and only if $\rho_{\bf A}$ contain non-classical correlations 
\cite{Piani-PRL106-2011, Streltsov-PRL106-2011}. In Figure \ref{Fig:APNCC}, we depict the activation protocol for three qubits.

We found that multiplicative negativity across ${\bf A}$ and ${\bf \tilde{A}}$ split is given as
\begin{eqnarray}
\mathcal{M} (\rho_{A:\tilde{A}})(\alpha) = \max \bigg[ \, 1\, , \, \frac{8 + 3 \, \alpha}{8} \, \bigg]\,, 
\end{eqnarray}
which is strictly larger than $1$ for any $\alpha > 0$. Hence, it follows that fully separable region of $\rho_3(\alpha)$ contain some non-classical 
correlations. As DQC1 model works for any $\alpha > 0$, so it follows that these non-classical correlations are responsible 
for the speed up of this quantum computational model and entanglement is not necessary for its operation. We recognize such 
correlations as {\it quantum dissonance} which can be present in fully separable states and quantum discord for correlations 
which may be quantified by measuring geometrical distance from any entangled state to a classically correlated state \cite{Modi-PRL104-2010}. 
   
\section{Summary}\label{conclusion}

We have revisited the problem of deterministic quantum computation with single qubit. We have provided evidence that there are instances where 
the presence of non-classical correlations, namely, quantum dissonance, in the DQC1 model is sufficient to provide quantum advantage over best 
known classical computation even when there is no entanglement of any kind anywhere in the setup. This example does not prove that there can be 
no bound entanglement for higher number of qubits rather just establishes the fact that this particular model of quantum computation can work 
without entanglement. We have analyzed an example for three qubits state which is the simplest multipartite system. We have shown that the final 
state of DQC1 model for three qubits can be written as a convex combination of a fully separable state and another state whose PPT region is fully 
separable. We have also indirectly demonstrated the existence of non-classical correlations via activating them into distillable entanglement. 
Further, we note that the first experiment to realize power of DQC1 was performed with only one qubit in the register \cite{Lanyon-PRL101-2008}. 
We believe that the same experiment could also be repeated for at least two qubits in the quantum register, which would be reasonably interesting. 

\ack
The author would like to thank Otfried G\"uhne for his generous and kind hospitality at University of Siegen. The author also acknowledges 
discussions with Otfried G\"uhne, Tobias Moroder, and S\"onke Niekamp. The author is thankful to referee for his comments.

\end{document}